\documentclass[aps,prl,twocolumn,tightenlines,superscriptaddress]{revtex4}
\usepackage{epsfig}
\usepackage{graphicx}
\usepackage{psfrag}
\usepackage{amsmath,amssymb,psfrag,slashed,graphicx}
\usepackage{colordvi}
\usepackage{color}
\def \t {\tilde}

\begin{document}

\title{{  Next-to-Next-to-Leading Order Calculation of Quasi Parton Distribution Functions}}

\author{Long-Bin Chen}
\affiliation{School of Physics and Materials Science, Guangzhou University, Guangzhou 510006, China}
\author{Wei Wang\footnote{Corresponding author:wei.wang@sjtu.edu.cn}}
\affiliation{INPAC,  SKLPPC, MOE KLPPC, School of Physics and Astronomy, Shanghai Jiao Tong University, Shanghai, 200240,   China}
\author{Ruilin Zhu\footnote{Corresponding author:rlzhu@njnu.edu.cn}}
\affiliation{Department of Physics and Institute of Theoretical Physics,
Nanjing Normal University, Nanjing, Jiangsu 210023, China}
\affiliation{Nuclear Science Division, Lawrence Berkeley National
Laboratory, Berkeley, CA 94720, USA}
\date{\today}
\vspace{0.5in}

\begin{abstract}
We present the  next-to-next-to-leading order (NNLO)  calculation of  quark quasi parton distribution functions (PDFs)  in the large momentum effective theory.  The nontrivial factorization at this order is established  explicitly and the full analytic matching coefficients between the   quasi distribution and the lightcone distribution  are derived. We   demonstrate that the NNLO numerical  contributions can   improve the behavior  of the extracted PDFs sizably.
With the unprecedented precision study of nucleon tomography at the planned electron-ion collider, high precision Lattice QCD simulations with our NNLO results implemented will enable to test the QCD theory and  more precise results on the PDFs of nucleons will be obtained.
\end{abstract}

\maketitle

{\it    Introduction.} The Feynman parton distribution functions (PDFs) are the most-important  cornerstones for applying quantum chromadynamics (QCD) to high energy particle and nuclear physics.
They provide  not only an important platform to unveil the fundamental structure of the nucleons, but are also a crucial ingredient to explore new physics beyond the standard model at hadron colliders.
Decades of extensive studies are made   to probe the  PDFs from hard QCD processes~\cite{Harland-Lang:2014zoa,Ball:2017nwa,Gao:2017yyd,Hou:2019efy}, while a limited success was achieved from the first principle of QCD, i.e., the Lattice QCD, and only a few lowest  moments were obtained~\cite{Martinelli:1987zd,Martinelli:1988xs,Detmold:2001dv,Dolgov:2002zm,Alexandrou:2019ali}.

Recently the large momentum effective theory (LaMET)~\cite{Ji:2013dva,Ji:2014gla}, established to calculate various parton distribution functions directly from lattice QCD,  has attracted great attentions from both phenomenology and lattice communities. Significant progress has been made, see, e.g., recent reviews~\cite{Cichy:2018mum,Ji:2020ect} and other applications~\cite{Radyushkin:2017cyf,Ma:2014jla,Ma:2017pxb}. In LaMET, a quasi-distribution is  constructed from the lattice calculable matrix element of hadron state and the relevant light-cone distributions can be derived through a perturbative matching. This provides a powerful tool to calculate all parton observables from the first principle of QCD which can be directly confronted with the experimental measurements. With  the unprecedented precision study of nucleon tomography at the planned electron-ion collider (EIC)~\cite{Accardi:2012qut}, high precision LaMET applications will enable us to test the QCD theory and  deepen  our understanding of  PDFs of nucleon.

According to the LaMET factorization, the quasi-PDF can be expressed in terms of lightcone-PDF,
\begin{align}\label{eq:fformula}
\tilde f_{i/H}(y ,p^z)=&\int_{-1}^1\frac{dx}{|x|}\Big[C_{ij}\Big(\frac{y}{x}, \frac{|x| p^z}{\mu} \Big)f_{j/H}(x, \mu) \Big] \ ,
\end{align}
where $\tilde f_{i/H}$ and $f_{j/H}$ represent the quasi-PDF and lightone-PDF, respectively, $i, j$ for the parton flavors and $\mu$ the factorization scale. In the above equation, $x\in [-1,1]$ and $y\in [-\infty, \infty]$ are the light-cone momentum and $\hat z$-component momentum fractions of the hadron carried by the parton $j$ and $i$, respectively. This factorization argument is obtained  on the basis that the Infrared (IR) behaviors for the quasi-PDF and lightcone-PDF are the same in LaMET~\cite{Ji:2013dva,Ji:2014gla}, and the matching coefficient $C_{ij}$ is perturbative calculable.

The fixed-order calculation plays an important role in the development of LaMET. It provides not only the explicit expression of the matching coefficients needed for the lattice computation, but also the detailed instances showing how the   factorization works. All previous analyses are based on  one-loop calculations~\cite{Ji:2020ect}. Very recently, it started to get into two-loop order, but only the ultraviolet (UV) renormalization was discussed in Ref.~\cite{Braun:2020ymy}. In this Letter, we will carry out, for the first time, the flavor non-singlet quark distribution in LaMET at two-loop order, including the matching coefficient and the numeric improvement to extract the lightcone-PDF from Lattice QCD.

We emphasize two important features of our study below. First, we will demonstrate the nontrivial feature of the QCD factorization at the NNLO. Soft divergences will be cancelled out between various contributions, whereas the collinear divergences between the quasi- and lightcone-PDFs cancel out. This cancellation requires the fine details of the theory, including $\epsilon$-term and the exact scale dependence in the one-loop matching. Our explicit demonstration provides an important proof of the factorization argument in LaMET~\cite{Green:2017xeu,Ji:2017oey,Ishikawa:2017faj,Li:2016amo}.

Second, the NNLO matching results can be directly implemented  in lattice calculations. As an example, we will show how this improves previous determination of the quark PDF in LaMET. This will have a significant impact in hadron physics community and will open new opportunities to perform high precision lattice PDF calculations in the new era~\cite{Lin:2020rut}.

To be explicit we will first derive the analytic  result for the flavor non-singlet quark distribution in LaMET at two-loop order, where   various techniques  developed for high-order calculations~\cite{Furmanski:1980cm,Curci:1980uw,Moch:2004pa,Vogt:2004mw} are employed. After subtracting the UV and IR divergences, we will derive the NNLO matching coefficient into two often-used renormalization schemes.  We will then show a numeric example where our new result can greatly improve the shape of the extracted quark PDFs.

\textit{LaMET factorization at two-loop order.}
We will focus on the flavor non-singlet quark distribution
whose light-cone distribution follows the usual definition in the literature,
\begin{align}\label{eq:pdf}
f_{q/H}(x,\mu)&=\!\int\!\! \frac{d\xi^-}{4\pi} \, e^{-ixp^+\xi^-}
\! \big\langle p \big| \bar{q}(\xi^-) \gamma^+ W(\xi^-,0)
q(0) \big|p\big\rangle,
\end{align}
where $W(\xi^-,0)$ denotes the lightcone gauge link. 
The quark quasi-distribution is defined as
\begin{align}\label{eq:pdf0}
{ \tilde  f}_{q/H}(y, p^z) &= \frac{p^z}{p^0} \int \frac{dz}{4\pi} e^{iz  yp^z}  \langle p|\overline{q} (z)
  \gamma^0 W(z,0) q(0) |p\rangle,~
\end{align}
where the Wilson link is along the $  z$ direction: $W(z,0)={\cal P}\exp\left(-ig \int_0^{z}dz' A^z(z') \right)$.

In the LaMET factorization of Eq.~(\ref{eq:fformula}), both $f_q$ and $\tilde{f}_q$ contain collinear divergences. The dimensional regulation with $D=4-2\epsilon$ and the minimal subtraction scheme ($\overline{\rm MS})$ are adopted in the calculations.
should be expanded as
\begin{align}
\t f^{(2)}_{i/k}(y, \frac{ p^z}{\mu},\epsilon_{\mathrm{IR}})=&C^{(2)}_{ij}\Big(\frac{y}{x}, \frac{|x| p^z}{\mu} \Big)\otimes f^{(0)}_{j/k}(x, \epsilon_{\mathrm{IR}})\nonumber\\&+C^{(1)}_{ij}\Big(\frac{y}{x}, \frac{|x| p^z}{\mu} \Big)\otimes f^{(1)}_{j/k}(x, \epsilon_{\mathrm{IR}})\nonumber\\&+C^{(0)}_{ij}\Big(\frac{y}{x}, \frac{|x| p^z}{\mu} \Big)\otimes f^{(2)}_{j/k}(x, \epsilon_{\mathrm{IR}}),\label{eq:matchNNLO}
\end{align}
where $\epsilon_{\rm IR}$ was introduced to regulate the collinear divergence and the convolution $\otimes$ integral is defined as in Eq.~(\ref{eq:fformula}). The perturbative expansion series are collected as $T_a=\sum_{n=0}^\infty \left(\frac{\alpha_s}{2\pi}\right)^n T_a^{(n)}$ with $T_a$ being each of $\t f_{i/k}, C_{ij}, f_{j/k}$. For the flavor non-singlet quark distribution, the collinear divergences in the lightcone PDFs $f_{j/k}^{(i)}$ on the right hand side of Eq.~(\ref{eq:matchNNLO}) are known in the literature~\cite{Furmanski:1980cm,Curci:1980uw,Moch:2004pa,Vogt:2004mw}. For the matching coefficients, the leading order is trivial:
$C^{(0)}_{ij}(y)=\delta_{ij}\delta(1-y)$, and the NLO $C^{(1)}_{ij}(y)$ in $\overline{\rm MS}$ and RI/MOM schemes can also be found in Refs.~\cite{Izubuchi:2018srq,Wang:2019tgg}.

\begin{figure}[th]
\begin{center}
\includegraphics[width=0.45\textwidth]{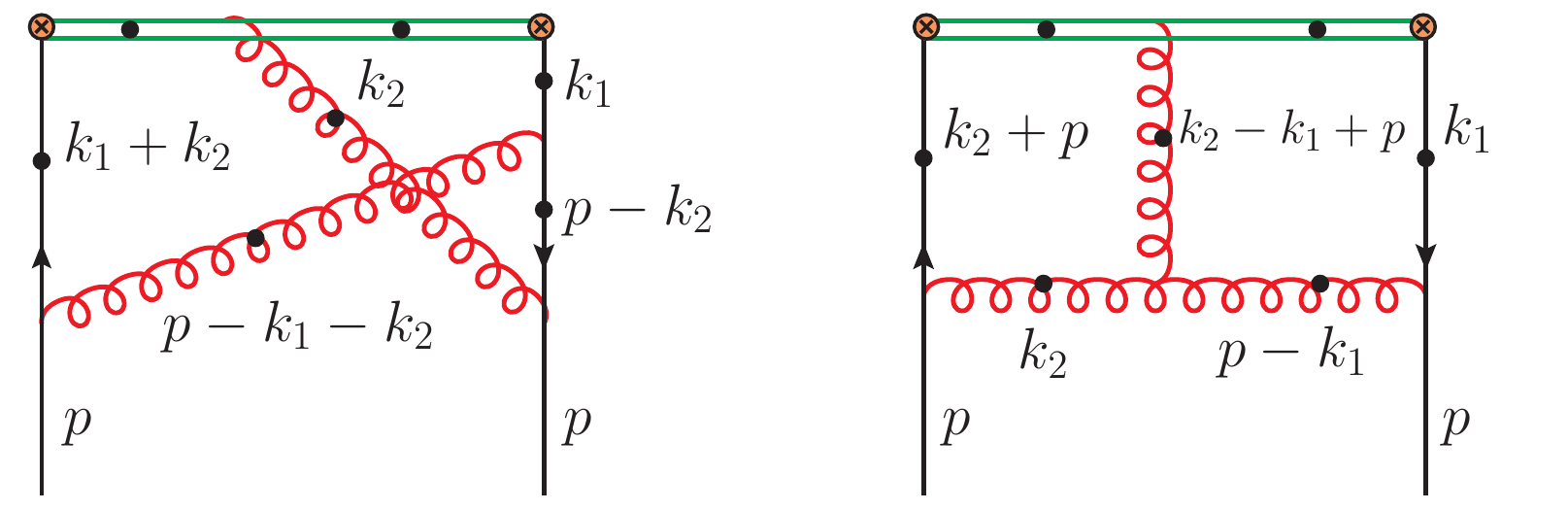}
\caption{ Feynman diagrams for the two-loop master integrals, where the double-lines correspond to  Wilson line. A dot on a propagator indicates that the power of the propagator is not always be $1$ and may be any integer $n_i$. }\label{fig:feynmandiagMI}
\end{center}
\end{figure}

Therefore, in order to demonstrate the factorization at NNLO, one needs to carry out the perturbative calculation of $\tilde f_q$ at two-loop order.
In total, there are 79 Feynman diagrams, and three
representative diagrams are shown in Fig.~\ref{fig:feynmandiagMI}. Virtual and real sub-diagrams can be obtained by applying  different cuts on the Wilson line and the total contributions satisfy  vector current conservation. We shall point out that all the Feynman integrals can be classified into three families of integrals:
\begin{align}
I^1_{n_i}&=\int\int \frac{d^D k_1 \, d^D k_2}{(k_{1}^{2})^{n_1}(k_{2}^{2})^{n_2}((k_{2}-p)^{2})^{n_3}((k_{1}+k_{2})^{2})^{n_4}}\nonumber \\ & \times \frac{1}{((k_{1}+k_{2}-p)^{2})^{n_5}} (\frac{1}{(P_1+i0)^{n_6}}-\frac{1}{(P_1-i0)^{n_6}})\nonumber \\ & \times\frac{1}{4\pi i} (\frac{1}{(Q_1+i0)^{n_7}}+\frac{1}{(Q_1-i0)^{n_7}}),
\end{align}
with
$P_1=n\cdot k_1+y n\cdot p$,
$Q_{1}=n\cdot k_2$  and $n^\mu=(0,\vec{0}_{D-2},1)$;
\begin{align}
I^2_{n_i}&=\int\int \frac{d^D k_1 \, d^D k_2}{(k_{1}^{2})^{n_1}(k_{2}^{2})^{n_2}((k_{2}-p)^{2})^{n_3}((k_{1}+k_{2})^{2})^{n_4}}\nonumber \\ & \times \frac{1}{((k_{1}+k_{2}-p)^{2})^{n_5}} (\frac{1}{(P_2+i0)^{n_6}}-\frac{1}{(P_2-i0)^{n_6}})\nonumber \\ & \times\frac{1}{4\pi i} (\frac{1}{(Q_2+i0)^{n_7}}+\frac{1}{(Q_2-i0)^{n_7}}),
\end{align}
with
$P_{2}=n\cdot k_1+n\cdot k_2+y n\cdot p$
and $Q_{2}=n\cdot k_2$;
\begin{align}
I^3_{n_i}&=\int\int \frac{d^D k_1 \, d^D k_2}{(k_{1}^{2})^{n_1}(k_{2}^{2})^{n_2}((k_{1}-p)^{2})^{n_3}((k_{2}+p)^{2})^{n_4}}\nonumber \\ & \times \frac{1}{((k_2-k_1+p)^2)^{n_7}} (\frac{1}{(P_3+i0)^{n_6}}-\frac{1}{(P_3-i0)^{n_6}})\nonumber \\ & \times\frac{1}{4\pi i} (\frac{1}{(Q_3+i0)^{n_5}}+\frac{1}{(Q_3-i0)^{n_5}}),
\end{align}
with
$P_{3}=n\cdot k_1+y n\cdot p$ and
$Q_{3}=n\cdot(k_2-k_1+p)$.
The $+i\epsilon$ prescription for the other propagators involving the $k_1, k_2$ are implicitly assumed.
The first two families of integrals $I^1_{n_i}$ and $I^2_{n_i}$  correspond to the two kinds of cut in the left diagram in Fig.~\ref{fig:feynmandiagMI}. The third family of integrals $I^3_{n_i}$ can be obtained by the right diagram in Fig.~\ref{fig:feynmandiagMI}.
To organize the calculations of these diagrams, we use FeynRules \cite{Alloul:2013bka} and FeynArts \cite{Hahn:2000kx}. The algebraic manipulation and simplification of the amplitudes are performed by  Mathematica packages FeynCalc \cite{Mertig:1990an}.
We employ  the integration-by-parts (IBP) techniques with the help of  FIRE \cite{Smirnov:2014hma} and reduce all the involved  tensor integrals  into a minimal set of integrals that are called master integrals (MIs).
We calculate all the MIs for both $p^2=0$ and $p^2\neq0$ cases with the method of differential equations~\cite{Kotikov:1990kg}. Inspired by Ref.~\cite{Henn:2013pwa}, we construct 3 groups of canonical basis $({\bf g}^i;i=1,2,3)$ that are linear combinations of MIs ~\cite{Chen:2020arf,Chen:2020iqi}, and whose differential equations can be expressed as:
\begin{eqnarray}
{\rm d}~{\bf g}^i =\epsilon\,~{\rm d}\text{\rm \bf M}\, \cdot {\bf g}^i,
\end{eqnarray}
{\rm \bf M} is matrix whose elements contain only log functions with rational coefficient, the above form will vastly simplify the calculations. More details on the calculations of MIs can be found in Ref.~\cite{Chen:2020iqi}.
These techniques developed  in this calculation are also applicable to other distributions including flavor-singlet quark and gluon PDFs and generalized parton distributions.

For the UV divergences,  the renormalization of the quasi-operator is given as
\begin{equation}
\t f(y, \frac{ p^z}{\mu},\epsilon_{\mathrm{IR}})=\int \frac{d y_1}{|y_1|}\left[\tilde{Z}\left(\frac{y}{y_1}\right)\right]\left[Z^{-1} \t f\left(y_1, \frac{ p^z}{\mu},\epsilon\right)\right],
\end{equation}
where $Z$ is the quark field wave function renormalization constant and $\tilde{Z}$ is the quasi
distribution renormalization constant~\cite{Ji:2015jwa,Braun:2020ymy}.
After subtracting the UV divergences, we are left with IR divergences. It contains $1/\epsilon_{\mathrm{IR}}$ and $1/{\epsilon_{\mathrm{IR}}^2}$ collinear divergences, and  can be expressed as
\begin{equation}\label{eq:div12}
\t f_{q/q}^{(2)}(y,\frac{p^z}{\mu},\epsilon_{\mathrm{IR}})|_{\mathrm{div}}=\frac{\Gamma^{\mathrm{IR}}_2(y)}{\epsilon_{\mathrm{IR}}^2}
+\frac{\Gamma^{\mathrm{IR}}_1(y)+2\Gamma^{\mathrm{IR}}_2(y)
\log(\frac{\mu^2}{{p^z}^{2}})}{\epsilon_{\mathrm{IR}}}.
\end{equation}
The explicit expressions for $\Gamma^{\mathrm{IR}}_2(y)$ and $\Gamma^{\mathrm{IR}}_1(y)$ are listed in the Supplemental material to this Letter~\cite{sup.mat.}. The $1/{\epsilon_{\mathrm{IR}}^2}$  divergence is cancelled by the last term of Eq.~(\ref{eq:matchNNLO}), whereas that of $1/\epsilon_{\mathrm{IR}}$ by the last two terms. At NNLO, these divergences depend on three color structures: $C_F^2$, $C_FC_A$ and $C_FT_F$. The cancellations of divergences are found for all these color structures. It is necessary to  emphasize that the explicit scale dependence in the one-loop matching plays an important role to demonstrate the complete cancellation of the collinear divergence.

\textit{Matching at NNLO.}
With the collinear divergence cancelled out completely in Eq.~(\ref{eq:matchNNLO}), one can  derive the matching coefficient at NNLO.
{  In the factorization formulae of Eqs.~(\ref{eq:fformula}) and (\ref{eq:matchNNLO}), the light-cone PDF is
defined  in the ${ \overline {\rm MS}}$ scheme while  the matching coefficient depends on the
 renormalization scheme of quasi-PDF.  The regularization-independent momentum subtraction (RI/MOM) scheme is mostly adopted in lattice calculations~\cite{Martinelli:1994ty}, while in some quasi-PDF studies, a two-step matching procedure has been advocated in Refs.\cite{Constantinou:2017sej,Alexandrou:2018pbm,Izubuchi:2018srq,Alexandrou:2018eet,Alexandrou:2019lfo}. An example is the so-called modified ${ \overline {\rm MS}}$ (${\rm M}\overline{\rm MS}$) renormalization scheme~\cite{Alexandrou:2019lfo}. In this scheme,  the lattice data on quasi-PDF is firstly converted to the ${\rm M}\overline{\rm MS}$ scheme, and in the second step one matches the ${\rm M}\overline{\rm MS}$-renormalized  quasi-PDF to the lightcone PDF.  Very recently a hybrid renormalization scheme has also been proposed in Ref.~\cite{Ji:2020brr}. }

The subtraction in RI/MOM scheme can be summarized as~\cite{Wang:2019tgg,Martinelli:1994ty}
\begin{align}
&\left.\tilde{Z}_{{ \rm{RI/MOM}}}^{-1} \left\langle p \left|\bar{q}(z) \gamma^{z} W(z, 0) q(0)\right| p \right\rangle\right|_{p^{2}
=-\mu_{R}^{2},p^z=p^z_{R}}\nonumber\\& =\left.\left\langle p \left|\bar{q}(z) \gamma^{z} W(z, 0) q(0)\right| p \right\rangle\right|_{\mathrm{LO}},
\end{align}
where $\mu_{R}$ and $p^z_{R}$ are the two renormalization scales in RI/MOM scheme.
The corresponding matching coefficient can be written as
\begin{eqnarray}\label{RIMOM}
C^{(n),{ \rm{RI/MOM}}}_{qq}&=& \bigg[C^{(n),\overline{\rm MS}}_{qq}\bigg(y, \frac{p^z}{\mu}\bigg)- (\tilde f_{q/q}^{(n)})_{C.T.}\bigg]_+,
\end{eqnarray}
where  $\left[C^{(n),\overline{\rm MS}}_{qq}\left(y,p^z/\mu\right)\right]_+$ is the n-th order matching coefficients in $\overline{\rm MS}$ scheme. The counter-term in the RI/MOM scheme is given by
\begin{eqnarray}\label{counterterm}
(\tilde f_{q/q}^{(n)})_{C.T.}= \left|\frac{p^z}{p^z_{R}}\right|  \tilde f_{q/q}^{(n),R}\left(\frac{p^z}{p^z_{R}}(y-1)+1, \frac{\mu_{R}^2}{{p^{z}_{R}}^2}\right).
\end{eqnarray}
{  The explicit expressions for these counter-terms are available  in the supplementary Mathematica package files
 to this Letter~\cite{sup.mat.}.}

With the factorization scale $\mu=p_z$, the matching coefficients in $\overline{\rm MS}$ scheme can be decomposed into three different color structures,
\begin{equation}
    C^{(2),\overline{\rm MS}}_{qq}(y,1)_{[i]}= \left(C_F {c}_i^{C_F}
   +C_A c_i^{C_A}+ 2T_F n_f c_i^{T_F}\right)C_F\ ,
\end{equation}
where $[i]$ represents four different kinematic regions for $y$: $y>1$, $0<y<1$, $-1<y<0$ and $y<-1$. One interesting point is that the scale dependent single logarithm $\log(\mu^2/{p^z}^2)$ appears in the NNLO matching coefficients at  all nonphysical regions. The complete expressions for $C^{(2),\overline{\rm MS}}_{qq}$ for all these regions are given in the Supplemental material to this Letter~\cite{sup.mat.}. Substituting the above results into Eq.~(\ref{RIMOM}), one can obtain the matching coefficients in the RI/MOM scheme.

In the ${\rm M}\overline{\rm MS}$ scheme, the matching coefficient is obtained from $C^{(2),{\rm M}\overline{\rm MS}}_{qq}(y,p^z/\mu)$ with  the asymptotic form in the $y\to \pm\infty$ region subtracted. All the expression of the matching coefficient  $C^{(2),{\rm M}\overline{\rm MS}}_{qq}(y,p^z/\mu)$ can be found in the supplemental material to this Letter~\cite{sup.mat.}.

\begin{figure}[th]
\begin{center}
\includegraphics[width=0.45\textwidth]{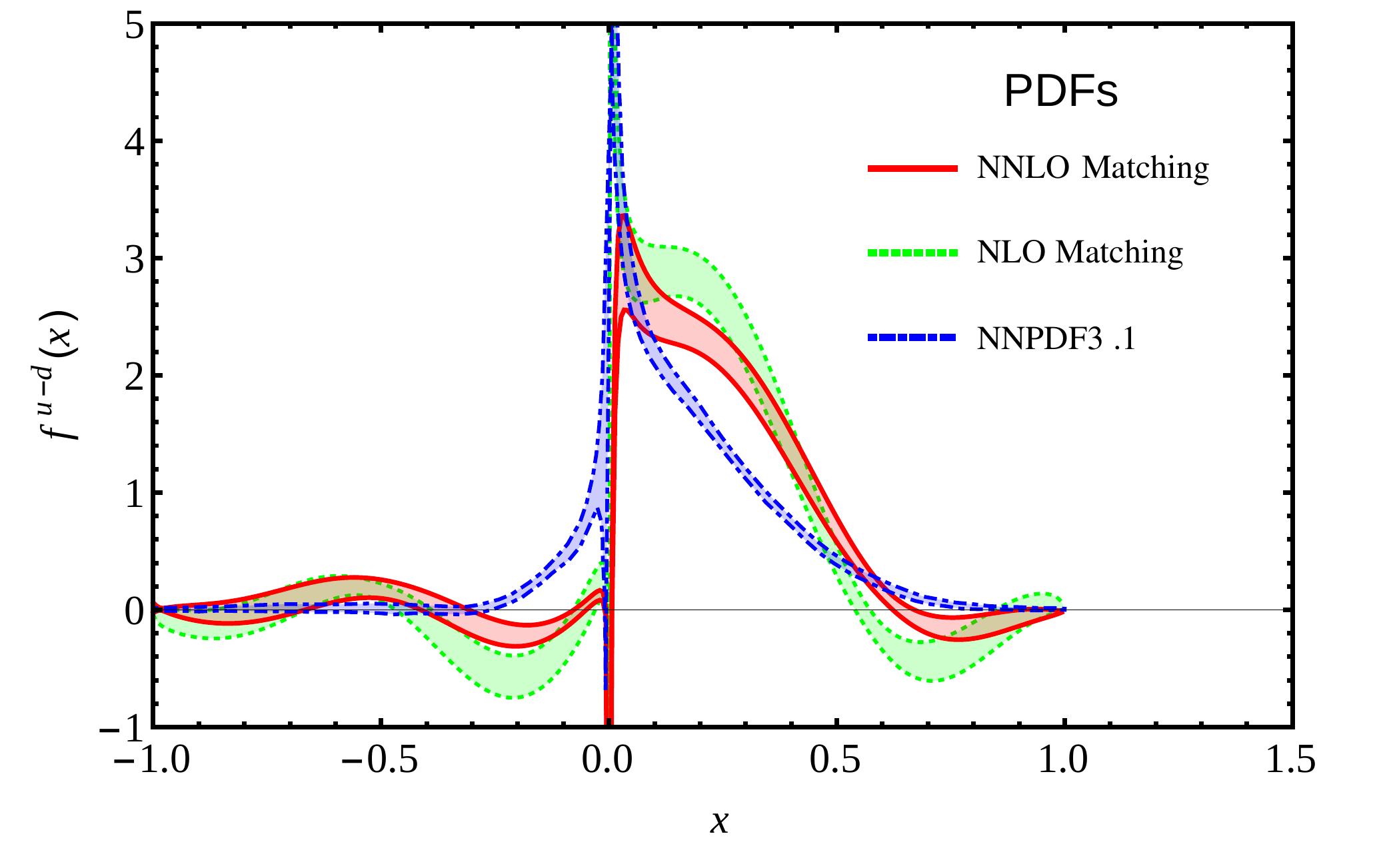}
\caption{ {  Results for the lightcone PDFs $f^{u-d}(x)$ at $\mu=2$GeV  using  the  lattice data in ${\rm M}\overline{\rm MS}$ renormalization scheme~\cite{Alexandrou:2019lfo}.
The result from the NNPDF3.1 global fit~\cite{Ball:2017nwa} is also shown as a comparison. An oscillatory behavior appears at NLO  due to the fact that the  lattice data has been  truncated  at $z=10a$ in coordinate space~\cite{Alexandrou:2019lfo,Liu:2018uuj}.} }\label{fig:pdf}
\end{center}
\end{figure}

\textit{Numerical Impact.}
{  We adopt here the ${\rm M}\overline{\rm MS}$ renormalization scheme to demonstrate the impact of NNLO results. We use the lattice data for the  ${\rm M}\overline{\rm MS}$-renormalized quasi-PDF from Ref.~\cite{Alexandrou:2019lfo}. As an example, in Fig.~\ref{fig:pdf}, we give the results of iso-vector quark distribution $f^{u-d}(x)$ extracted from the lattice data of~\cite{Alexandrou:2019lfo} at NLO and NNLO, respectively. } In the numeric calculations, we choose $\mu=2$GeV and $p^z=2.3$GeV. One can see from Fig.~\ref{fig:pdf} that the NNLO correction is important to improve the NLO behavior and the extracted distribution at large $x$ region agrees better with the phenomenology fit from the NNPDF3.1 set~\cite{Ball:2017nwa}.  An oscillatory behavior appears  because  the cut-off method is used and we truncate the lattice data at $z=10a$ in coordinate space~\cite{Lin:2017ani,Liu:2018uuj,Alexandrou:2019lfo}. The NNLO corrections can
soften the oscillatory behavior.  We plan to have a more  detailed comparison of different schemes and a detailed analysis of theoretical uncertainties in a future publication.

\textit{Conclusions.} We have  for the first time explored the flavor-non-singlet quark quasi-PDFs in the large momentum effective theory at two-loop order. With the explicit full analytic results, we found that all the collinear divergences  factorized into the relevant lightcone PDFs. This has provided a concrete proof of the LaMET factorization at the nontrivial two-loop order. The matching coefficient between the quark quasi-PDF and lightcone-PDF was derived in the ${ \overline {\rm MS}}$ and RI/MOM subtraction scheme. As an example, we have also shown in the ${\rm M}\overline{\rm MS}$ scheme  that the NNLO corrections improve the previous lattice result for the iso-vector quark distribution.

We expect that more theoretical developments   will follow along the direction of this Letter. In particular, the procedure and computation techniques can be extended to all other channels, including flavor singlet quark distribution and gluon distribution functions. This will complete all necessary ingredients for extracting PDFs from lattice QCD at two-loop order. Our calculation can be applied to other parton observables, such as the generalized parton distributions,  transverse momentum dependent distributions and meson distribution amplitudes. This will provide a solid ground for applying lattice QCD to nucleon tomography and comparing to the experiment exploration from the future EIC.

\textit{Acknowledgements}---We thank F. Yuan for all the valuable advices and discussions during the work. We thank X. Ji, Y. Ji, H.-n. Li, Y.-S. Liu,  J. Wang, J. Xu,  L.-L. Yang, S. Zhao, Y. Zhao, J.-H. Zhang, Q.-A. Zhang for valuable discussions, in particular S. Zhao for discussions  on the RI/MOM subtraction. We appreciate F. Yuan and X. Ji for reading and polishing the manuscript. {  We thank Z.Y. Li and Y.Q. Ma for the help in the comparison of our results with theirs. }  
LBC is supported by the National Natural Science Foundation of China (NSFC) under the grant No.~11805042. WW is supported by NSFC under grants No.~11735010, 11911530088,  by Natural Science Foundation of Shanghai under grant No. 15DZ2272100. RLZ is supported by NSFC under grant No.~11705092, by Natural Science Foundation of Jiangsu under Grant No.~BK20171471 and Jiangsu Qing Lan Project, by China Scholarship Council under Grant No.~201906865014 and partially supported by the U.S. Department of Energy, Office of Science, Office of Nuclear Physics, under contract number DE-AC02-05CH11231. 
The numerical calculation is supported by  the $\pi$ 2.0 cluster supported by the Center for High Performance Computing at Shanghai Jiao Tong University. 

\textit{Note Added}---When this manuscript is being prepared, a preprint~\cite{2006.12370} appears, in which the authors calculated the two-loop corrections to  the onshell quark correlation functions defined in the coordinate space, and the results are in agreement  with ours.

\begin{widetext}
\section*{Supplemental material }

In this supplemental material, we give the explicit analytic expressions for the divergences of quasi PDFs and the NNLO matching coefficients calculated in the main text.

\subsection{Divergences}

The quasi PDFs have both the UV and IR divergences.  The $1/{\epsilon^2}$  divergence appears only in physical  $y$ region $0<y<1$ and the  expression for its coefficient $\Gamma_2(y)$ is
\begin{align}
\Gamma_2(y)=&\theta(y)\theta(1-y)\left[\frac{C_{F}^{2}}{ 2}\left(\frac{\left(3 y^2+1\right) \log (y)-4 \left(y^2+1\right) \log (1-y)+2 (y-1)^2}{y-1}\right)+\frac{C_{F}\beta_0\left(y^2+1\right)}{2 (y-1) }\right],
\end{align}
where $\beta_0=\frac{11 C_A-4T_F n_f}{6}=\frac{11 C_A-2 n_f}{6}$. Therein the expression for the  coefficient  of $1/\epsilon_{\mathrm{IR}}^2$ IR divergence, i.e., $\Gamma^{\mathrm{IR}}_2(y)$ defined in Eq.~(9) of the main text, can be written as
\begin{align}
\Gamma^{\mathrm{IR}}_2(y)=&\Gamma_2(y)-\left(\beta_0+\frac{3}{2}C_F\right)\frac{1+y^2}{y-1}C_F\theta(y)\theta(1-y).
\end{align}

Because the $1/\epsilon$ divergences parts as well as the finite terms do not vanish outside $0<y<1$, we have four different kinematic regions for $y$: $y>1$, $0<y<1$, $-1<y<0$ and  $y<-1$.
The $1/\epsilon$  divergences   appear in all the y region and expressions for $\Gamma_1(y)$  are

\begin{align}
\Gamma_1(y)|_{y>1}=&C_F^2\bigg[-\frac{\left(y^2+3\right) \text{Li}_2(1-y)}{y-1}+\frac{\pi ^2 \left(y^2+3\right)-18 (y-1)}{6-6 y}-\frac{2
   \left(y^2+1\right) \log ^2(y-1)}{y-1}-\frac{\left(3 y^2+1\right) \log ^2(y)}{2 (y-1)}\nonumber\\&+\left(\frac{\left(3 y^2+1\right)
   \log (y)}{y-1}+2 (y-1)\right) \log (y-1)-2 (y-1) \log (y)\bigg]+C_F\frac{\beta_0 \left(\left(y^2+1\right) \log \left(\frac{y-1}{y}\right)+y-1\right)}{ (y-1)},\\
   \Gamma_1(y)|_{0<y<1}=& C_{F}^2\bigg[\frac{\left(3 y^2+1\right) \text{Li}_2(y)}{y-1}+\frac{4 \left(y^2+1\right) \log ^2(1-y)}{y-1}-\frac{\left(17 y^2+7\right)
   \log ^2(y)}{4 (y-1)}+\frac{\left(4 y^2+10
   y-9\right) \log (y)}{2 (y-1)}\nonumber\\&+\frac{2 \log (2) \left(4 \left(y^2+1\right) \log (1-y)-\left(3 y^2+1\right) \log (y)-2
   (y-1)^2\right)}{y-1}+\frac{3 (y+1)}{2}\nonumber\\&+\frac{\left(-6 y^2+3 \left(y^2+1\right) \log (y)+4 y-6\right) \log (1-y)}{y-1}\bigg]+ \bigg[-\frac{11 \left(y^2+1\right) \log (1-y)}{6 (y-1)}-\frac{11 \left(y^2+1\right) \log
   (2)}{3 (y-1)}\nonumber\\&+\frac{253 y^2-3 \pi ^2 \left(y^2+1\right)-174 y+187}{36 (y-1)}+\frac{\left(y^2+1\right) \log ^2(y)}{4 (y-1)}-\frac{\left(17
   y^2+5\right) \log (y)}{12 (y-1)}\bigg]C_A C_{F}\nonumber\\&~~+2C_FT_F n_f \bigg[-\frac{17 y^2-6 y+11}{18 (y-1)}+\frac{\left(y^2+1\right) \log (1-y)}{3 (y-1)}+\frac{\left(y^2+1\right) \log (y)}{6
   (y-1)}+\frac{2 \left(y^2+1\right) \log (2)}{3 (y-1)}\bigg],
   \end{align}
   \begin{align}
   \Gamma_1(y)|_{-1<y<0}=& C_{F}^2\bigg[(y+1) \text{Li}_2(y)+\frac{2 \left(y^2+1\right) \log ^2(1-y)}{y-1}+\frac{y^2 \log ^2(-y)}{y-1}-\frac{2
   \left(\left(y^2+1\right) \log (-y)+(y-1)^2\right) \log (1-y)}{y-1}\nonumber\\&+2 y+\frac{\pi ^2}{3-3 y}+(y-1) \log (-y)-1\bigg]+ C_A C_{F}\bigg[-\frac{\left(y^2+1\right) \text{Li}_2(y)}{y-1}-\frac{\left(12+\pi ^2\right) y^2+22 y+\pi ^2-34}{12
   (y-1)}\nonumber\\&+\frac{\left(y^2+1\right) \log ^2(-y)}{4 (y-1)}+\frac{\left(7 y^2-3 y+7\right) \log (-y)}{3
   (y-1)}-\frac{\left(y^2+1\right) \log (1-y) (6 \log (-y)+11)}{6 (y-1)}\bigg]\nonumber\\&~~+2C_FT_Fn_f \bigg[\frac{\left(\left(y^2+1\right) \log (1-y)-\left(y^2+1\right) \log (-y)+y-1\right)}{3 (y-1)}\bigg],\\
   \Gamma_1(y)|_{y<-1}=&-\Gamma_1(y)|_{y>1}.
\end{align}
The expression for the  coefficient  of $1/\epsilon_{\mathrm{IR}}$ IR divergence, i.e. $\Gamma^{\mathrm{IR}}_1(y)$ in Eq.~(9) of the main text, can be written as
\begin{align}
\Gamma^{\mathrm{IR}}_1(y)=&\Gamma_1(y)-\left(\beta_0+\frac{3}{2}C_F\right)C^{(1),\overline{\rm MS}}_{qq}\left(y,\frac{p^z}{\mu}\right),
\end{align}
where the expressions of $C^{(1),\overline{\rm MS}}_{qq}\left(y,\frac{p^z}{\mu}\right)$ will be given in the following section.

\subsection{NNLO matching coefficients}

The explicit expressions for the  counter-terms in RI/MOM scheme are very tedious and  available to download from the supplementary Mathematica package files
 to both the Letter and the arXiv version. The explicit analytic expressions  of NNLO matching coefficients in ${\rm M}\overline{\rm MS}$ scheme can be written as
\begin{align}
C^{(2),{\rm M}\overline{\rm MS}}_{qq}(y, \frac{p^z}{\mu})|_{y>1}= & C^{(2),\overline{\rm MS}}_{qq}(y, \frac{p^z}{\mu})|_{y>1}+\frac{C_F  \left(11 C_A+9 C_F-2 n_f\right)}{4 y}\log \left(\frac{\mu ^2}{{p^z}^2}\right)-\frac{C_F n_f
   (5-4 \log (2 y))}{4 y}\nonumber\\&-\frac{C_A C_F \left(132 \log (2 y)+4 \pi ^2-159\right)}{24 y}+\frac{C_F^2 \left(-108 \log (2 y)+16 \pi ^2+75\right)}{24 y},\\
   C^{(2),{\rm M}\overline{\rm MS}}_{qq}(y, \frac{p^z}{\mu})|_{0<y<1}=&  C^{(2),\overline{\rm MS}}_{qq}(y, \frac{p^z}{\mu})|_{0<y<1},\\
   C^{(2),{\rm M}\overline{\rm MS}}_{qq}(y, \frac{p^z}{\mu})|_{-1<y<0}=& C^{(2),\overline{\rm MS}}_{qq}(y, \frac{p^z}{\mu})|_{-1<y<0},\\
   C^{(2),{\rm M}\overline{\rm MS}}_{qq}(y, \frac{p^z}{\mu})|_{y<-1}= &C^{(2),\overline{\rm MS}}_{qq}(y, \frac{p^z}{\mu})|_{y<-1}+\frac{C_F  \left(11 C_A+9 C_F-2 n_f\right)}{4 (1-y)}\log \left(\frac{\mu ^2}{{p^z}^2}\right)-\frac{C_F n_f
   (5-4 \log (-2 y))}{4(1- y)}\nonumber\\&-\frac{C_A C_F \left(132 \log (-2 y)+4 \pi ^2-159\right)}{24 (1-y)}+\frac{C_F^2 \left(-108 \log (-2 y)+16 \pi ^2+75\right)}{24 (1-y)}.
 \end{align}
The explicit analytic expressions of  NNLO matching coefficients in $\overline{\rm MS}$ scheme are
\begin{align}
C^{(2),\overline{\rm MS}}_{qq}(y, \frac{p^z}{\mu})|_{y>1}= & \left(C_F {c}_1^{C_F}
   +C_A c_1^{C_A}+2T_Fn_f c_1^{T_F}\right)C_F+ \left(\Gamma_1(y)|_{y>1}\right)\log (\frac{\mu^2}{p_z^2}),\\
   C^{(2),\overline{\rm MS}}_{qq}(y, \frac{p^z}{\mu})|_{0<y<1}=& C_F \left(C_F {c}_2^{C_F}
   +C_A c_2^{C_A}+2T_Fn_f c_2^{T_F}\right)+\left({\Gamma}_2(y)\right)\log^2 (\frac{\mu^2}{p_z^2})\nonumber\\&+ \left(\left(\Gamma_1(y)|_{0<y<1}\right) -(P^{(1),V}_{qq}(y)|_{0<y<1})\right)\log (\frac{\mu^2}{p_z^2}),\\
   C^{(2),\overline{\rm MS}}_{qq}(y, \frac{p^z}{\mu})|_{-1<y<0}=& C_F \left(C_F {c}_3^{C_F}
   +C_A {c}_3^{C_A}+2T_Fn_f c_3^{T_F}\right)+ \left(\left(\Gamma_1(y)|_{-1<y<0}\right) -P^{(1),V}_{\bar{q}q}(-y)\right)\log (\frac{\mu^2}{p_z^2}),
         \end{align}
   \begin{align}
   C^{(2),\overline{\rm MS}}_{qq}(y, \frac{p^z}{\mu})|_{y<-1}= &-C^{(2),\overline{\rm MS}}_{qq}(y, \frac{p^z}{\mu})|_{y>1}\ .
 \end{align}

 For convenience, we also list the explicit analytic expressions  of NLO matching coefficients in ${\rm M}\overline{\rm MS}$ scheme \cite{Alexandrou:2019lfo,Alexandrou:2018pbm}
\begin{align}
C^{(1),{\rm M}\overline{\rm MS}}_{qq}(y, \frac{p^z}{\mu})|_{y>1}= & C^{(1),\overline{\rm MS}}_{qq}(y, \frac{p^z}{\mu})|_{y>1}+\frac{3C_F}{2 y},\\
   C^{(1),{\rm M}\overline{\rm MS}}_{qq}(y, \frac{p^z}{\mu})|_{0<y<1}=&  C^{(1),\overline{\rm MS}}_{qq}(y, \frac{p^z}{\mu})|_{0<y<1},\\
   C^{(1),{\rm M}\overline{\rm MS}}_{qq}(y, \frac{p^z}{\mu})|_{-1<y<0}=& C^{(1),\overline{\rm MS}}_{qq}(y, \frac{p^z}{\mu})|_{-1<y<0},\\
   C^{(1),{\rm M}\overline{\rm MS}}_{qq}(y, \frac{p^z}{\mu})|_{y<-1}= &C^{(1),\overline{\rm MS}}_{qq}(y, \frac{p^z}{\mu})|_{y<-1}+\frac{3C_F}{2 (1-y)}.
 \end{align}
The $C^{(1),\overline{\rm MS}}_{qq}\left(y,\frac{p^z}{\mu}\right)$ terms can also be found in~\cite{Izubuchi:2018srq,Wang:2019tgg} and a
complete discussion on the regularization at $y\to \infty$ at one-loop is investigated in~\cite{Izubuchi:2018srq}.
\begin{align}
C^{(1),\overline{\rm MS}}_{qq}\left(y,\frac{p^z}{\mu}\right)_{y>1}=& C_{F}
\left[\frac{1+y^{2}}{1-y} \ln \frac{y}{y-1}+1\right], \\
C^{(1),\overline{\rm MS}}_{qq}\left(y,\frac{p^z}{\mu}\right)_{0<y<1}=& C_F\left[\frac{1+y^{2}}{1-y}\left(-\ln \frac{\mu^{2}}{4p_{z}^{2}}+\ln ( y(1-y))\right)-\frac{y(1+y)}{1-y}\right],\\
C^{(1),\overline{\rm MS}}_{qq}\left(y,\frac{p^z}{\mu}\right)_{y<0}=& C_F\left[-\frac{1+y^{2}}{1-y} \ln \frac{y}{y-1}-1\right].
\end{align}

The two-loop valence to valence quark/antiquark splitting functions are given in Refs.~\cite{Moch:2004pa,Vogt:2004mw}
\begin{align}
P_{q\bar{q}}^{(1),V}(y)=& \frac{C_F}{2}\left(\frac{C_{A}}{2}-C_{F}\right)\left[\frac{1+y^{2}}{1+y}\left(\ln ^{2}(y)
-4 \mathrm{Li}_{2}(-y)-4 \ln (y+1) \ln (y) -\frac{\pi^{2}}{3}\right)+4(1-y)+2(y+1) \ln (y)\right].\\
P_{qq}^{(1),V}(y)=&\frac{1}{2} C_{F}^{2}\left[\left(6 \zeta_{3}+\frac{3}{8}-\frac{\pi^{2}}{2}\right) \delta(1-y)-\frac{1}{2}(1+y) \ln ^{2}(y)+\left(\frac{7}{2} y^{2}-2 y-\frac{3}{2}\right)\frac{ \ln (y)}{1-y}-5(1-y)\right. \nonumber\\
&\left.-\left(2\ln (1-y)+\frac{3}{2}\right) \left(1+y^{2}\right)\ln (y) \left[\frac{1}{1-y}\right]_+\right]+\frac{1}{2} C_{A} C_{F}\left[\left(\frac{67}{9}-\frac{\pi^{2}}{3}\right) \left[\frac{1}{1-y}\right]_+
\right.\nonumber\\
&\left.+\left(-3 \zeta_{3}+\frac{17}{24}+\frac{11 \pi^{2}}{18}\right) \delta(1-y) +\frac{\left(1+y^{2}\right) \ln ^{2}(y)}{2(1-y)}+\frac{\left(5 y^{2}+17\right) \ln (y)}{6(1-y)}+\frac{53-187 y}{18}+(1+y)\frac{\pi^{2}}{6}\right] \nonumber\\
&-\frac{1}{2} C_{F} T_{F} n_{f}\left[\frac{20}{9} \left[\frac{1}{1-y}\right]_++\left(\frac{1}{6}+\frac{2 \pi^{2}}{9}\right) \delta(1-y)+\frac{2\left(1+y^{2}\right) \ln (y)}{3(1-y)}+\frac{4(1-y)}{3}-\frac{10(y+1)}{9}\right].
\end{align}

In the following, we will give the explicit expressions of ${c}_i^{C_F}$,
  $c_i^{C_A}$ and $c_i^{T_F}$. The $c_i^{C_F}$ auxiliary functions are
\begin{align}
{c}_1^{C_F}&=\frac{4 \left(2 y^2+1\right) \text{Li}_3\left(\frac{1}{1-y}\right)}{y-1}+\frac{\left(3 y^2+1\right)
   \text{Li}_3\left(\frac{1}{y}\right)}{y-1}-\frac{2 \left(y^2+1\right)
   \left(\text{Li}_3\left(1-y^2\right)+\text{Li}_3\left(-\frac{1}{y}\right)-2
   \text{Li}_3\left(\frac{1}{y+1}\right)\right)}{y-1}\nonumber\\&+\frac{\text{Li}_2\left(\frac{1}{y}\right)
   \left(y^2 (1+2 \log (2))-2 \left(y^2-1\right) \log (y-1)+4 \left(y^2+1\right) \log (y+1)-8+6
   \log (2)\right)}{y-1}+2 (y-1) \text{Li}_2\left(-\frac{1}{y}\right)\nonumber\\&+\frac{2 \left(2 y^2+1\right) \log ^3(y)}{3 (y-1)}-\frac{2
   \left(y^2+1\right) \log ^3(y+1)}{3 (y-1)}+\frac{2 \log (2) \left(2 \left(y^2+1\right) \log
   ^2\left(\frac{y-1}{y}\right)+3 (1-y)+2 (y-1)^2 \log
   \left(\frac{y}{y-1}\right)\right)}{y-1}\nonumber\\&-\frac{\left(\pi ^2 \left(y^2+1\right)+6 \log (y) \left(y^2 \log (y)+2
   \left(y^2+1\right) (\log (y+1)-1)\right)+3 (y-1)^2\right) \log (y-1)}{3
   (y-1)}+\frac{3
   y-2}{y-1}+\frac{2 \pi ^2}{3-3 y}\nonumber\\&-\frac{\left(\left(12+\pi ^2\right) y^2+\pi ^2-12\right) \log (y+1)}{3 (y-1)}-\frac{2 (2 ((y-1) y+1)+\log (y)) \log ^2(y-1)}{y-1}\nonumber\\&+\frac{2 \log ^2(y) \left(\left(y^2+1\right) \log (y+1)-2
   y\right)}{y-1}+\frac{2 \left(y^2+2\right)
   \log ^3(y-1)}{3 (y-1)}+(5 y-3)
   \log (y),
 \end{align}

\begin{align}
{c}_2^{C_F}&=-\frac{2 \left(y^2+1\right) \left(\text{Li}_3\left(1-y^2\right)+7 \text{Li}_3(-y)+2
   \text{Li}_3\left(\frac{y}{y+1}\right)+3\zeta (3)\right)}{y-1}-\frac{\left(y^2+3\right)
   \text{Li}_3(y)}{y-1}-\frac{\left(y^2+5\right) \text{Li}_3\left(\frac{y}{y-1}\right)}{y-1}\nonumber\\&-\frac{\text{Li}_2(y) \left(-11 y^2+\left(3 y^2-1\right) \log
   (1-y)-\left(y^2+1\right) \log \left(\frac{y}{64 (y+1)^4}\right)+10 y+2-4 \log (2)\right)}{y-1}\nonumber\\&
   +\text{Li}_2(-y) \left(\frac{4
   \left(y^2+1\right) \log (y)}{y-1}+2 y+6\right)-\frac{\left(-30 y^2+\pi ^2 \left(y^2+3\right)+51 y-3\right) \log (y)}{6 (y-1)}+\frac{\pi ^2 \left(y \left(-3 y^2+y+2\right)-1\right)}{3 (y-1)}\nonumber\\&-\frac{\left(11 y^2+7\right) \log
   ^3(1-y)}{6 (y-1)}+\frac{7 \left(5 y^2+3\right) \log ^3(y)}{12 (y-1)}-\frac{\left(3 \left(y^2+1\right) \log (y)+y ((y-7) y+5)-7\right) \log ^2(1-y)}{y-1}\nonumber\\&+\frac{\left(-7 y^2+3
   \left(y^2+1\right) \log ^2(y)+2 \log (y) \left(-4 \left(y^2+1\right) \log (y+1)+y (y (2 y+5)-4)+1\right)+y-2\right)
   \log (1-y)}{2 (y-1)}\nonumber\\&-\frac{2 \left(y^2+1\right) \log (y) \log ^2(y+1)}{y-1}+\frac{2 \log ^2(2) \left(-4
   \left(y^2+1\right) \log (1-y)+3 y^2 \log (y)+2 (y-1)^2+\log (y)\right)}{y-1}\nonumber\\&+\frac{2 \left(y^2+1\right) \log ^3(y+1)}{3
   (y-1)}+\frac{2 \log (2) \left(\left(7
   y^2+2\right) \log ^2(y)-2 \left(y^2+1\right) \log (1-y) (2 \log (1-y)+\log (y))\right)}{y-1}\nonumber\\&+\frac{3 y^2-16 y+11}{2-2 y}+\frac{2 \log (2) \left(y^2+9 y^2 \log (1-y)+\left(-3 y^2-4
   y+6\right) \log (-(y-1) y)-5 y+4\right)}{y-1}\nonumber\\&+\frac{\left(\left(\pi ^2-12\right) y^2+12 \left(y^2-1\right) \log (y)+\pi
   ^2+12\right) \log (y+1)}{3 (y-1)}-\frac{(2 y (2 y (y+8+6 \log (2))+1)-25) \log ^2(y)}{4 (y-1)},
 \end{align}

\begin{align}
{c}_3^{C_F}&=-\frac{\left(5 y^2+7\right) \text{Li}_3(y)}{y-1}-\frac{2 \left(y^2+1\right) \left(\text{Li}_3(-y)+2
   \text{Li}_3\left(\frac{2 y}{y-1}\right)-2 \text{Li}_3\left(\frac{y}{y+1}\right)+2
   \text{Li}_3\left(\frac{2 y}{y+1}\right)+2
   \text{Li}_3\left(\frac{y+1}{y-1}\right)\right)}{y-1}\nonumber\\&+\frac{4 y^2
   \text{Li}_3\left(\frac{y}{y-1}\right)}{y-1}-\frac{\text{Li}_2(y) \left(2
   \left(y^2-1\right) \log (1-y)-2 \left(y^2+1\right) \log \left(-\frac{y}{8 (y+1)^2}\right)+(4-11 y)
   y+6-4 \log (2)\right)}{y-1}\nonumber\\&+\frac{2 \text{Li}_2(-y)
   \left(\left(y^2+1\right) \log (-y)+2 (y-1)\right)}{y-1}+\frac{-11 y^2+6 y+7}{2-2 y}+\frac{\pi ^2
   \left(4 y^2-2 y-1\right)}{6 (y-1)}-\frac{2 \left(2 y^2+1\right) \log ^3(1-y)}{3
   (y-1)}\nonumber\\&+\frac{2 \left(-\left(y^2+1\right) \log
   (y+1)+2 (y-1) y+2\right) \log ^2(1-y)}{y-1}-\frac{2 \left(y^2+1\right) \log (2) \left(2 \log
   ^2(1-y)+\log ^2\left(\frac{y+1}{1-y}\right)\right)}{y-1}\nonumber\\&-\frac{4 y^2 \zeta (3)}{y-1}-\frac{\left(y^2-2\right) \log ^3(-y)}{3 (y-1)}+\frac{\left(\pi ^2 \left(y^2+1\right)+3
   (y-1)^2+6 (3 y+1) (y-1) \log (-y)\right) \log (1-y)}{3 (y-1)}\nonumber\\&-\frac{\left(3 \left(7+\pi ^2\right)
   y^2-42 y+\pi ^2+9\right) \log (-y)}{6 (y-1)}-\frac{\left(\left(12+\pi ^2\right) y^2-6
   \left(y^2-1\right) \log (-y)+\pi ^2-12\right) \log (y+1)}{3 (y-1)}\nonumber\\&+\frac{((4-5 y) y-2) \log
   ^2(-y)}{2 (y-1)}-\frac{1}{3} \log (2) \left(\pi ^2 y+24 y+3 (y+1) \log ^2(-y)-12 (y-1) \log
   (1-y)+\pi ^2+6\right).
 \end{align}

The $c_i^{C_A}$ auxiliary functions are

\begin{align}
 c_1^{C_A}&=\frac{\left(y^2+1\right) \left(\text{Li}_3\left(1-y^2\right)-2
   \text{Li}_3\left(\frac{1}{1-y}\right)+\text{Li}_3\left(-\frac{1}{y}\right)+\text{Li}_3\left(\frac{1}{y}\right)-2
   \text{Li}_3\left(\frac{1}{y+1}\right)\right)}{y-1}+\frac{((87-128 y) y-59) \log (y)}{9
   (y-1)}\nonumber\\&+\frac{\text{Li}_2\left(\frac{1}{y}\right)
   \left(5 y^2+12 \left(y^2+1\right) \log \left(\frac{y}{y+1}\right)+17\right)}{6 (y-1)}-(y-1)
   \text{Li}_2\left(-\frac{1}{y}\right)+\frac{\left(y^2+1\right) \log ^3(y-1)}{3 (y-1)}+\frac{2
   \left(y^2+1\right) \log ^3(y)}{3 (y-1)}\nonumber\\&+\frac{\left(y^2+1\right) \log ^3(y+1)}{3 (y-1)}-\frac{11
   \left(y^2+1\right) \log ^2(y-1)}{6 (y-1)}+\frac{\left(10 (y (11 y-12)+11)-9 \left(y^2+1\right)
   \log ^2(y)\right) \log (y-1)}{9 (y-1)}\nonumber\\&-\frac{11
   \log (2) \left(\left(y^2+1\right) \log \left(\frac{y-1}{y}\right)+y-1\right)}{3
   (y-1)}+\frac{\left(2 \left(y^2-1\right)+\left(y^2+1\right) \log \left(\frac{(y-1)^2}{y}\right)
   \log (y)\right) \log (y+1)}{y-1}\nonumber\\&+\frac{11 \left(y^2+1\right) \log ^2(y)}{6 (y-1)}+\frac{\pi ^2 \left(\left(y^2+1\right) \log
   \left(y^2-1\right)+1\right)}{6 (y-1)}+\frac{4}{3-3 y}+\frac{151}{18},\\
 c_2^{C_A}&=\frac{\left(y^2+1\right) \left(\text{Li}_3\left(1-y^2\right)+7 \text{Li}_3(-y)+2 \text{Li}_3(y)-3
   \text{Li}_3\left(\frac{y}{y-1}\right)+2 \text{Li}_3\left(\frac{y}{y+1}\right)\right)}{y-1}+\frac{1261
   y^2-1119 y+880}{54 (y-1)}\nonumber\\&+\text{Li}_2(-y)
   \left(-\frac{2 \left(y^2+1\right) \log (y)}{y-1}-y-3\right)+\frac{\text{Li}_2(y) \left(-y^2+6 \left(y^2+1\right) \log
   \left(\frac{(1-y) (y+1)^2}{y^3}\right)+26\right)}{6 (y-1)}-\frac{5 \left(y^2+1\right) \zeta (3)}{2 (y-1)}\nonumber\\&+\frac{\left(y^2+1\right) \log ^3(1-y)}{2 (y-1)}-\frac{ \left(y^2+1\right) (7\log ^3(y)+4\log ^3(y+1))}{12
   (y-1)}-\frac{\left(-19 y^2+6 \left(y^2+1\right) \log (y)-6 y-19\right)
   \log ^2(1-y)}{12 (y-1)}\nonumber\\&-\frac{\left(193 y^2+18 \left(y^2+1\right) \log ^2(y)+9 \left(y^2+2 y-3\right) \log (y)-204
   y+211\right) \log (1-y)}{18 (y-1)}+\frac{\left(67 y^2-24 y-35\right) \log ^2(y)}{24 (y-1)}\nonumber\\&+\frac{\log (2) \left(3 \pi ^2 y^2-220 y^2-9 \left(y^2+1\right) \log ^2(y)+33
   \left(y^2+1\right) \log (1-y)+18 \left(y^2-1\right) \log (y)+207 y+3 \pi ^2-187\right)}{9 (y-1)}\nonumber\\&+\frac{11
   \left(y^2+1\right) \log ^2(2)}{3 (y-1)}-\frac{\left(161 y^2-183 y+89\right) \log (y)}{18 (y-1)}+\frac{\pi ^2
   \left(-y^2+6 \left(y^2+1\right) \log \left(\frac{(1-y)^3 y}{y+1}\right)+12 y+2\right)}{36 (y-1)}\nonumber\\&+\frac{\left(y^2+1\right)
   \log (y) \log ^2(y+1)}{y-1}+\frac{2
   \left(y^2+\left(-y^2+\left(y^2+1\right) \log (1-y)+1\right) \log (y)-1\right) \log (y+1)}{y-1},\\
 {c}_3^{C_A}&=\frac{\left(y^2+1\right) \left(\text{Li}_3(-y)+2 \text{Li}_3(y)+2 \text{Li}_3\left(\frac{2
   y}{y-1}\right)-2 \text{Li}_3\left(\frac{y}{y+1}\right)+2 \text{Li}_3\left(\frac{2 y}{y+1}\right)+2
   \text{Li}_3\left(\frac{y+1}{y-1}\right)\right)}{y-1}\nonumber\\&-\frac{\text{Li}_2(-y)
   \left(\left(y^2+1\right) \log (-y)+2 (y-1)\right)}{y-1}+\frac{\text{Li}_2(y) \left(-25 y^2+6
   \left(y^2+1\right) \log \left(-16 y (y+1)^2\right)+12 y+11\right)}{6 (y-1)}\nonumber\\&+\frac{-99 y^2-302
   y+449}{36 (y-1)}-\frac{2 \left(y^2+1\right) \log ^3(1-y)}{3 (y-1)}-\frac{7 \left(y^2+1\right) \log
   ^3(-y)}{12 (y-1)}+\frac{\left(y^2+1\right) (6 \log (-y)+11) \log ^2(1-y)}{6 (y-1)}\nonumber\\&+\frac{\left(-20
   \left(11 y^2-12 y+11\right)+27 \left(y^2+1\right) \log ^2(-y)-18 \left(5 y^2-2 y+1\right) \log
   (-y)\right) \log (1-y)}{18 (y-1)}\nonumber\\&+\frac{\left(y^2+12 y-23\right) \log ^2(-y)}{12
   (y-1)}+\frac{\left(y^2+1\right) \log (2) \left(\log ^2\left(\frac{1-y}{y+1}\right)+\log
   \left(-\frac{(y-1)^4}{y}\right) \log (-y)\right)}{y-1}\nonumber\\&+\frac{\left(\left(y^2+1\right) \log
   ^2(1-y)-\left(y^2-1\right) (\log (-y)-2)\right) \log (y+1)}{y-1}+\frac{\left(485 y^2-330
   y+209\right) \log (-y)}{36 (y-1)}\nonumber\\&+\frac{\log (2) \left(\pi ^2 y^2+12 y^2+11 \left(y^2+1\right)
   \log (1-y)+\left(-17 y^2+12 y-17\right) \log (-y)+11 y+\pi ^2-23\right)}{3 (y-1)}\nonumber\\&+\frac{\pi ^2
   \left(-2 y^2+3 \left(y^2+1\right) \log (-y (y+1))+3 y+1\right)}{18 (y-1)}.
 \end{align}

The $c_i^{T_F}$ auxiliary functions are
\begin{align}
 c_1^{T_F}&=-\frac{\left(y^2+1\right) \text{Li}_2\left(\frac{1}{y}\right)}{3 (y-1)}+\frac{\left(y^2+1\right)
   \log ^2(y-1)}{3 (y-1)}-\frac{\left(y^2+1\right) \log ^2(y)}{3 (y-1)}-\frac{2 \left(7 y^2-6
   y+7\right) \log (y-1)}{9 (y-1)}\nonumber\\&+\frac{2 \log (2) \left(\left(y^2+1\right) \log
   \left(\frac{y-1}{y}\right)+y-1\right)}{3 (y-1)}+\frac{2 \left(7 y^2-3 y+4\right) \log (y)}{9
   (y-1)}+\frac{14-11 y}{9 (y-1)},\\
 c_2^{T_F}&=-\frac{\left(y^2+1\right) \text{Li}_2(y)}{3 (y-1)}-\frac{\pi ^2 \left(y^2+1\right)}{18 (y-1)}-\frac{151 y^2-84 y+103}{54
   (y-1)}-\frac{\left(y^2+1\right) \log ^2(1-y)}{3 (y-1)}+\frac{\left(y^2+1\right) \log ^2(y)}{12 (y-1)}\nonumber\\&-\frac{2
   \left(y^2+1\right) \log ^2(2)}{3 (y-1)}+\frac{2 \left(7 y^2-6 y+7\right) \log (1-y)}{9 (y-1)}-\frac{2 \log (2)
   \left(-14 y^2+3 \left(y^2+1\right) \log (1-y)+9 y-11\right)}{9 (y-1)}\nonumber\\&+\frac{\left(11 y^2-12 y+11\right) \log (y)}{18
   (y-1)},\\
 c_3^{T_F}&=-\frac{\left(y^2+1\right) \text{Li}_2(y)}{3 (y-1)}-\frac{\pi ^2 \left(y^2+1\right)}{18
   (y-1)}-\frac{\left(y^2+1\right) \log ^2(1-y)}{3 (y-1)}+\frac{\left(y^2+1\right) \log ^2(-y)}{6
   (y-1)}+\frac{2 \left(7 y^2-6 y+7\right) \log (1-y)}{9 (y-1)}\nonumber\\&-\frac{2 \left(7 y^2-3 y+4\right) \log
   (-y)}{9 (y-1)}-\frac{2 \log (2) \left(\left(y^2+1\right) \log (1-y)-\left(y^2+1\right) \log
   (-y)+y-1\right)}{3 (y-1)}+\frac{14-11 y}{9-9 y}.
 \end{align}

\end{widetext}

\end{document}